\documentclass[preprint,showkeys,superscriptaddress,amsmath,amssymb,aps,pra]{revtex4-1}

\usepackage{upgreek}
\usepackage{times}
\usepackage{graphicx}
\usepackage{dcolumn}
\usepackage{amsmath}
\usepackage[dvipsnames]{xcolor}
\usepackage{amssymb}
\usepackage[version=3]{mhchem}

\begin{document}

\newcommand{\BeginManuscript}{%
		\renewcommand{\figurename}{\textbf{Figure}}
     }
\BeginManuscript

\title{Increasing efficiency of high numerical aperture metasurfaces using the grating averaging technique}

\author{Amir Arbabi}
\email{arbabi@umass.edu}
\affiliation{Department of Electrical and Computer Engineering, University of Massachusetts Amherst, 151 Holdsworth Way, Amherst, MA 01003, USA}
\author{Ehsan Arbabi}
\affiliation{T. J. Watson Laboratory of Applied Physics, California Institute of Technology, 1200 E. California Blvd., Pasadena, CA 91125, USA}
\author{Mahdad Mansouree}
\affiliation{Department of Electrical and Computer Engineering, University of Massachusetts Amherst, 151 Holdsworth Way, Amherst, MA 01003, USA}
\author{Seunghoon Han}
\affiliation{Samsung Advanced Institute of Technology, Samsung Electronics, Samsung-ro 130, Suwon-si, Gyeonggi-do 443-803, South Korea}
\author{Seyedeh Mahsa Kamali}
\affiliation{T. J. Watson Laboratory of Applied Physics, California Institute of Technology, 1200 E. California Blvd., Pasadena, CA 91125, USA}
\author{Yu Horie}
\affiliation{T. J. Watson Laboratory of Applied Physics, California Institute of Technology, 1200 E. California Blvd., Pasadena, CA 91125, USA}
\author{Andrei Faraon}
\email{faraon@caltech.edu}
\affiliation{T. J. Watson Laboratory of Applied Physics, California Institute of Technology, 1200 E. California Blvd., Pasadena, CA 91125, USA}

\begin{abstract}
One of the important advantages of optical metasurfaces over conventional diffractive optical elements is their capability to efficiently deflect light by large angles. However, metasurfaces are conventionally designed using approaches that are optimal for small deflection angles and their performance for designing high numerical aperture devices is not well quantified. Here we introduce and apply a technique for the estimation of the efficiency of high numerical aperture metasurfaces. The technique is based on a particular coherent averaging of diffraction coefficients of periodic blazed gratings and can be used to compare the performance of different metasurface designs in implementing high numerical aperture devices. Unlike optimization-based methods that rely on full-wave simulations and are only practicable in designing small metasurfaces, the gradient averaging technique allows for the design of arbitrarily large metasurfaces. Using this technique, we identify an unconventional metasurface design and experimentally demonstrate a metalens with a numerical aperture of 0.78 and a measured focusing efficiency of 77\%. The grating averaging is a versatile technique applicable to many types of gradient metasurfaces, thus enabling highly efficient metasurface components and systems.
\end{abstract}

\maketitle

\section*{Introduction}
Flat optical devices based on dielectric metasurfaces have recently attracted significant attention due to their small size, low weight, and the potential for their low-cost manufacturing using semiconductor fabrication techniques~\cite{Kildishev2013a,Yu2014,Estakhri2016,Jahani2016,Lalanne2017,Staude2017,Hsiao2017,Kamali2018}. The planar form factor of metasurfaces and the high multilayer overlay accuracy of the semiconductor manufacturing process enable the implementation of low-cost monolithic optical systems composed of cascaded metasurfaces whose production does not involve post-fabrication assembly and alignment steps~\cite{Arbabi2016d,Faraji-Dana2018}. Used either as single-layer devices or as integral parts of cascaded metasurface systems, one of the main requirements for metasurfaces is high efficiency. As a result, increasing the efficiency of metasurfaces has been the subject of recent studies~\cite{Arbabi2015,Zheng2015,Decker2015,Byrnes2016,Arbabi2017,Sell2017,Sell2018,Pestourie2018,Mansouree2018,Campbell2019}, and low numerical aperture (NA) metasurface components (i.e., metasurfaces with small deflection angles) with efficiencies of more than 97\% have been reported~\cite{Arbabi2015h}. However,  the efficiency of metasurfaces is known to decrease with increasing NA, resulting in a trade-off between the NA and efficiency~\cite{Astilean1998,Lalanne1998,Lalanne1999,Arbabi2015}. Several approaches have been proposed for designing efficient high-NA metasurfaces including adjoint optimization~\cite{Sell2017,Mansouree2018,Pestourie2018,Campbell2019,Phan2019,Chung2019} and patching together separately designed gratings~\cite{Byrnes2016,Paniagua-Dominguez2018}. The metasurfaces designed based on iterative adjoint optimization techniques can be efficient~\cite{Sell2017,Mansouree2018,Pestourie2018,Campbell2019}; however, their  dimensions are inevitably limited by the available computational resources because the optimization process  requires full-wave simulation of the entire device in each iteration, or simulation and optimization of each zone of the device~\cite{Phan2019} that is applicable to metalenses but cannot be readily extended to general holograms. Separate optimization of small portions of a gradient metasurface and then assembling them together to form a large metalens has been proposed~\cite{Byrnes2016,Paniagua-Dominguez2018}; however, the effect of discontinuities at the boundaries of patches has not been considered and efficient high-NA metasurfaces based  on such approaches have not been demonstrated yet. 

Typical metasurfaces are arrays of scatterers (or meta-atoms) arranged on 2D lattices. The conventional approach of designing metasurfaces involves selection of a set of parameterized meta-atoms and finding  a one-to-one map between the desired optical response (e.g., phase shift) and the meta-atom parameters. The map typically involves one or two of the meta-atom parameters, and a number of other design parameters such as the meta-atom height, geometry, and lattice constant are selected by the designer. These parameters are usually selected for achieving high transmission (or reflection) and a $2\pi$ phase coverage, but their effects on the performance of high-NA metasurfaces designed using the same design maps cannot be easily evaluated. Here we introduce a novel approach for evaluating the performance of different metasurface designs in implementing high-NA metasurface components. The approach is based on the adiabatic approximation of aperiodic metasurfaces by periodic blazed gratings and considers the effect of large deflection angles. Using the proposed  approach, we identify a design for implementing efficient  high-NA metasurfaces and experimentally demonstrate a metalens  with an NA of 0.78 and a focusing efficiency of 77\% as well as a 50$^\circ$ beam deflector with more than 70\% deflection efficiency for unpolarized light.

We focus our study on the design of gradient metasurfaces composed of meta-atoms that are arranged on periodic 2D lattices. This category represents a large class of metasurfaces and has been used in the realization of different types of metasurface optical elements~\cite{Aieta2012,Farmahini2013,Ni2013a,Pors2013,Lin2014a,Yang2014,Arbabi2014,Wu2017} and systems~\cite{Arbabi2016d,Faraji-Dana2018}.  We first discuss the conventional technique for designing transmissive gradient metasurfaces and then use this discussion to explain the main idea of the proposed grating averaging technique. 

The conventional approach for designing transmissive gradient metasurfaces assumes a local complex transmission coefficient for each meta-atom that only depends on the meta-atom itself (i.e., it is independent of the neighboring meta-atoms)~\cite{Aieta2012,Farmahini2013,Ni2013a,Pors2013,Yang2014,Arbabi2015}. For metasurfaces that operate under normal incidence, the local transmission coefficient for each meta-atom is approximated by the transmission coefficient of a periodic array created by periodically arranging the same meta-atom on the metasurface lattice. A library of meta-atoms is typically formed that spans the range of required transmission coefficients (e.g., 0 to $2\pi$ phase shift range). A flat optical component with a desired spatially varying complex transmission coefficient $t(x,y)$ is realized by an aperiodic meta-atom array whose local transmission coefficients best approximate $t(x,y)$ at the location of the meta-atoms (i.e., the lattice sites). Figure 1 shows an example of the conventional approach for designing metasurfaces composed of nano-post meta-atoms~\cite{Arbabi2014,Vo2014,Arbabi2015}. In this case, the meta-atom library comprises square-cross-section nano-posts with different widths $W$ that should provide complex transmission coefficients  $\mathrm{exp}(-j\phi)$ for $\phi$ spanning the $0$ to $2\pi$ range.  Figure 1a shows the simulated transmission coefficient of periodic arrays of amorphous silicon nano-posts (height: 500 nm, width: 60 nm-260 nm,  lattice constant: 400 nm, wavelength: 915 nm) that are used as estimates of local transmission coefficients in designing aperiodic metasurfaces. Refractive indices of 3.65 and 1.45 are assumed for the amorphous silicon and the fused silica substrate, respectively. The inverted relation that maps the desired transmission coefficient $\mathrm{exp}(-j\phi)$, which is indexed by $\phi$, to the nano-posts width is shown in Fig. 1b. A metasurface implementing an arbitrary phase profile can be designed  by sampling the desired phase profile at each lattice site and  determining the nano-post's width at that location using the design curve shown in Fig. 1b. 

\begin{figure}[t!]
\centering
\includegraphics[width=\columnwidth]{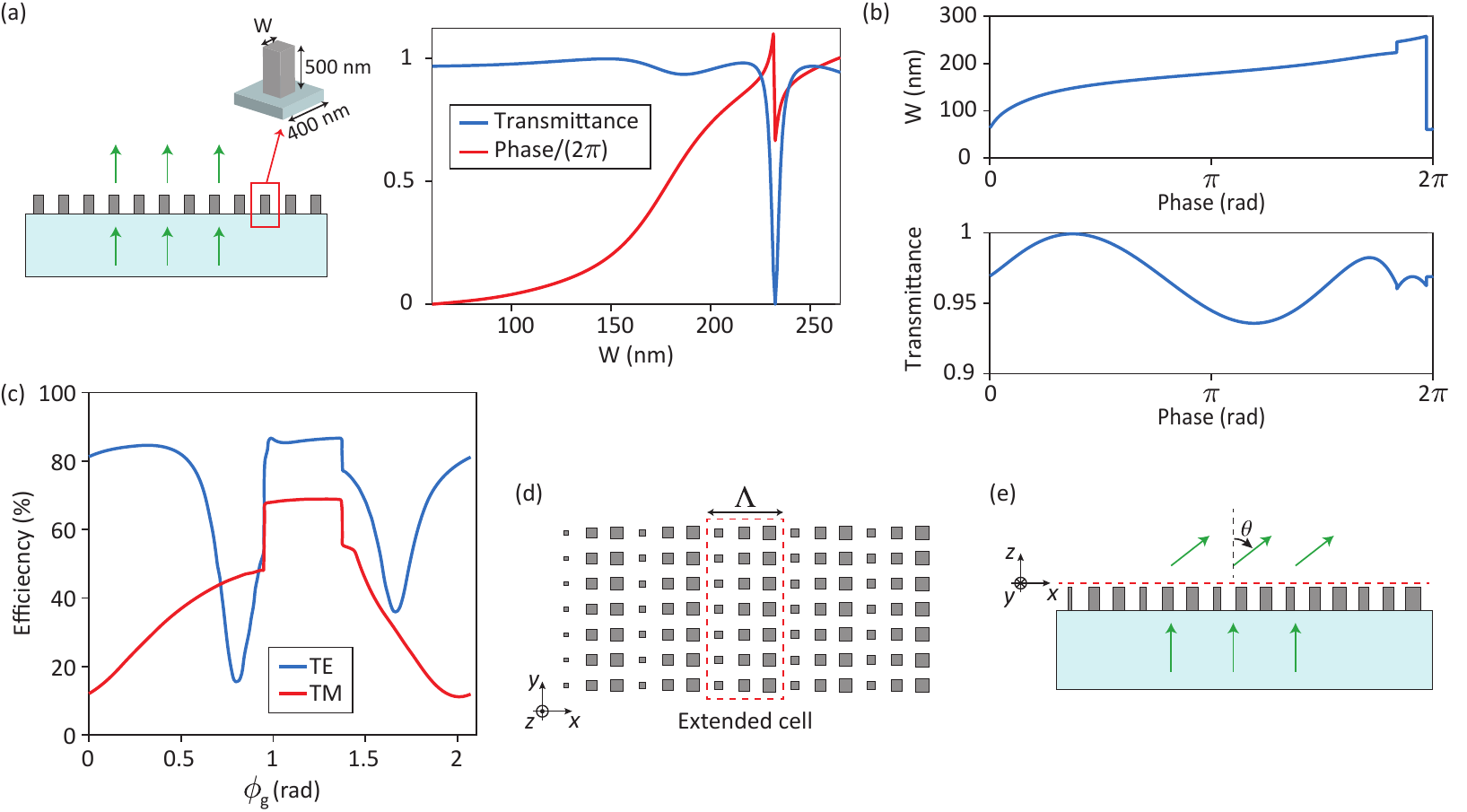}
\caption{(a) Illustration of an example periodic metasurface composed of amorphous silicon nano-posts with square cross-sections and its transmission coefficient for normally incident light as a function of the nano-post's width ($W$). (b) The design curve that maps desired phase shifts to nano-post widths (top plot) and the corresponding transmittance (bottom plot). The design curve is obtained from the transmission coefficient data shown in (a). (c) Diffraction efficiencies of the first transmitted order of blazed gratings with different phases ($\phi_\mathrm{g}$). The blazed gratings are designed using the metasurface platform and design curve shown in (a) and (b). The gratings' period is 1200 nm (i.e., three times the metasurface lattice constant) corresponding to a first order diffraction angle of 49.7$^\circ$. The efficiency values are computed for light normally incident from the substrate with transverse electric (TE) and  transverse magnetic (TM) polarizations.  (d) Top view of an aperiodic metasurface with a deflection angle of 52$^\circ$  that is designed using the metasurface platform and design curves shown in (a) and (b). The nano-post dimensions vary rapidly from one metasurface unit cell to the next, but slowly from one extended cell to the next.  (e) Side view of the aperiodic beam deflector shown in (d). The dashed red line depicts the output plane of the device. All results are for 915 nm light.}
\end{figure}

The conventional approach provides a simple and scalable technique for designing metasurfaces composed of a large number of meta-atoms. However, two main approximations are used in this approach.  First,  it assumes that the metasurface is locally periodic with a period equal to the lattice constant. In other words, it ignores the change in the coupling between meta-atoms and their neighbors that is caused by the aperiodicity of the metasurface. Second,  it assumes that the radiation pattern of the meta-atoms (i.e., the element factor) is isotropic, thus the local transmission coefficient of the meta-atoms is assumed to be independent of the incident and scattered directions. The first approximation is justified if the meta-atom shapes change gradually across the metasurface. For the validity of the second approximation, the radiation pattern of the meta-atoms should not change significantly over the angular range of interest for the incident and transmitted light. As a result, the efficiency of metasurfaces designed using the conventional approach depends on the coupling strength among the meta-atoms and the angular dependence of their radiation patterns.

Low-NA metasurfaces are composed of gradually varying meta-atoms and satisfy the requirement for the first approximation. They also deflect incident light by small angles, thus the condition for the second approximation is also satisfied when the incident angle is small.  As a result, low-NA metasurfaces designed using the conventional approach can be highly efficient and have achieved experimentally measured efficiency values as high as 97\% ~\cite{Arbabi2015h}. These conditions are also satisfied for the low-NA metasurfaces that operate under oblique incidence provided the design maps are also obtained for the same incident angles~\cite{Faraji-Dana2018}. For a more general case where the incident angle varies across a low-NA metasurface, one may use different design maps for different regions of the metasurface based on the local incident angle (i.e., phase gradient of the incident field).

As the metasurface NA increases, the approximations involved in the conventional design approach become less accurate and the metasurface efficiency decreases. Qualitatively, the performance of a metasurface platform in implementing high-NA devices designed using the conventional approach depends on the coupling among the meta-atoms and their radiation patterns. However, there is no fast approach to evaluate, compare, and predict the performance of different designs for the realization of high-NA metasurfaces. The grating averaging technique discussed in the next section addresses this issue.

\clearpage
\section*{Results}
\noindent\textbf{Grating averaging technique}

Beam deflectors are basic elements in designing gradient metasurfaces because such metasurfaces can be considered as beam deflectors with gradually varying deflection angles. As a result, the deflection efficiency of beam deflectors designed using a metasurface platform can be used to evaluate the performance of the platform in realizing general metasurface components.  Metasurface beam deflectors implement a linear phase ramp (i.e., $t=\mathrm{exp}(-j\phi(x,y))$ where $\phi(x,y)$ is a linear function of $x$ and $y$) and can be designed for arbitrary deflection angles. A beam deflector in the $z=0$ plane  that deflects normally incident light propagating along the $z$ direction by an angle $\theta$ toward the $x$ axis has a phase profile of $\phi(x)=2\pi/\lambda\sin(\theta)x+\phi_0$, where $\lambda$ is the light's wavelength in the $z>0$ region and $\phi_0$ is a constant. Now, consider implementing such a beam deflector by wrapping its phase to 0--2$\pi$ range and using a metasurface with a square lattice with the lattice constant of $a$. For specific values of $\theta$,  $a\sin(\theta)/\lambda$ is a rational number (i.e., $a\sin(\theta)/\lambda=n/m,$ where $n$ and $m$ are coprime integers) and the implemented metasurface beam deflector is periodic along the $x$ direction. The period is $\Lambda=ma$ and the beam deflector may be considered as an $n^\mathrm{th}$-order blazed grating. When $a\sin(\theta)/\lambda$ is an irrational number the metasurface is aperiodic and we refer to it as an aperiodic beam deflector.  One might consider approximating the local diffraction efficiency of aperiodic beam deflectors by the diffraction efficiency of a periodic grating with approximately the same deflection angle. This is particularly interesting because the diffraction efficiency of gratings can be computed using fast computational methods such as the rigorous coupled mode analysis (RCWA) technique~\cite{Moharam1981}. However, there are two issues that need to be addressed regarding this approximation. 

First, for a given grating period and a design map, there is a family of blazed gratings with the same deflection angle but different phases and efficiencies. The phase profile of an ideal blazed grating that deflects normally incident light by an angle $\theta_\mathrm{g}$ is given by $\phi(x)=2\pi/\lambda\sin(\theta_\mathrm{g})x+\phi_\mathrm{g}$ where $\phi_\mathrm{g}$ is a constant representing the phase of the diffracted light at $x=0$. Different values for $\phi_\mathrm{g}$ lead to different blazed grating designs with different efficiencies. For example, using the metasurface design curve shown in Fig. 1b, three-post blazed gratings can be designed by setting the phase delays imparted by the  three nano-posts as $\phi_\mathrm{g}$, $\phi_\mathrm{g}+2\pi/3$ and $\phi_\mathrm{g}+4\pi/3$. Different three-post blazed gratings that are obtained for different values of $\phi_\mathrm{g}$  have the same period of $3a=1200$ nm and diffraction angle of $\theta_\mathrm{g}=\sin^{-1}(\lambda/(3a))=49.7^\circ$ when illuminated with a normally incident 915 nm light from the substrate side. The simulated diffraction efficiencies of these gratings (for the +1 diffraction order) as a function of $\phi_\mathrm{g}$ for two incident polarizations are shown in Fig. 1c. As  Fig. 1c shows, the diffraction efficiency of the gratings varies significantly with $\phi_\mathrm{g}$; therefore, the deflection efficiency of a periodic metasurface beam deflector is not unique and depends on its phase. 

Now consider an aperiodic beam deflector with a deflection angle $\theta$ close to the diffraction angle of a blazed grating $\theta_\mathrm{g}$. For large deflection angles, the meta-atoms vary significantly from one lattice site (unit cell) to the next along the direction of the phase gradient. However, the aperiodic beam deflector can be considered as a slowly varying blazed grating. This can be seen in Fig. 1d that shows the top view of a portion of an aperiodic beam deflector with the deflection angle of $\theta=52^\circ$ that is designed using the design map shown in Fig. 1b. The meta-atoms in the beam deflector shown in Fig. 1d vary rapidly from one unit cell to the next, but slowly between extended cells containing three meta-atoms. Therefore, each extended cell of the aperiodic beam deflector may be approximately considered as a period of a blazed grating with some value of $\phi_\mathrm{g}$, and from one extended cell to the next, the value of $\phi_\mathrm{g}$  varies slowly. As a result, it is reasonable to estimate the deflection efficiency of aperiodic beam deflectors using the diffraction response of blazed gratings with approximately the same diffraction angle. 

Consider an aperiodic beam deflector as schematically shown in Fig. 1e and assume that the beam deflector is composed of a large number of extended cells. The beam deflector deflects a normally incident plane wave by an angle $\theta$ which is close to the diffraction angle $\theta_\mathrm{g}$ of a family of blazed gratings with different phases  $\phi_\mathrm{g}$.  Depending on the polarization of the incident wave, the deflected light is either TE  or TM polarized with respect to $z$. As shown in the Supplementary Note 1, the deflection coefficient of the aperiodic beam deflector for either TE or TM polarization is given by
\begin{equation}
A\approx\frac{1}{2\pi}\int_0^{2\pi}{t_n(\phi_\mathrm{g})e^{j\phi_\mathrm{g}}\mathrm{d}\phi_\mathrm{g}},
\end{equation}
where $t_n(\phi_\mathrm{g})$ represents the diffraction coefficient for the same polarization of a blazed grating with the diffraction angle $\theta_\mathrm{g}$ designed with the phase of $\phi_\mathrm{g}$. The phase of the diffraction coefficient is the same as the phase of the electric field of the diffracted wave at $x=z=0$ and its amplitude is given by $|t_n(\phi_\mathrm{g})|=\sqrt{\eta_{g}}$, where $\eta_\mathrm{g}$ is the diffraction efficiency of the blazed grating.  The deflection efficiency (i.e., the ratio of the power of the deflected beam and the incident beam power) for the aperiodic grating is given by $\eta=|A|^2$.  

The deflection efficiency of aperiodic beam deflectors can be obtained according to (1) which is a specially weighted average of the complex-valued diffraction coefficients of blazed gratings with the same diffraction angle and different phases. We can compute the deflection efficiency of an aperiodic beam deflector by designing $N$ different blazed gratings with different $\phi_{\mathrm{g}_i}=\frac{2\pi}{N},\frac{4\pi}{N},..., 2\pi$, finding their diffraction coefficients $t_n(\phi_{\mathrm{g}_i})$ and approximating  the integral in (1) by  $\frac{1}{N}\sum_{i=1}^N{t_n(\phi_{\mathrm{g}_i})e^{j\phi_{\mathrm{g}_i}}}$. The main advantage is that the diffraction coefficients of the blazed gratings can be computed quickly.

In the ideal case, the diffraction coefficient of a blazed grating designed for the phase of $\phi_\mathrm{g}$ is $t_n(\phi_\mathrm{g})=\exp(-j\phi_\mathrm{g})$ leading to $\eta=|A|^2=1$ (according to (1)) for the ideal beam deflector. In practice, the diffraction efficiency of the designed gratings (i.e.,  $|t_\mathrm{g}|^2$) is smaller than unity and there is a difference between their actual and desired phases $\phi_\mathrm{g}$. Both of these will lead to the reduction of the efficiency of aperiodic beam deflectors.  

In contrast to the periodic beam deflectors (i.e., blazed gratings), the efficiencies of aperiodic beam deflectors are well-defined and  independent of their phases. For example, the deflection efficiency of a beam deflector with the deflection angle of $\theta_\mathrm{g}=\sin^{-1}(\lambda/(3a))=49.7^\circ$ which is designed using the metasurface platform of Fig. 1 may be any of the values shown in Fig. 1c; however, the deflection efficiency of a large beam deflector with the deflection angle of $50^\circ$ which is designed using the same design curve is uniquely obtained from (1) and is $\sim$66\% and $\sim$32\% for the TE and TM polarizations, respectively. 

Low-NA metasurfaces are a special case where the extended cell is the same as the metasurface unit cell, grating diffraction angle is zero ($\theta_\mathrm{g}=0$), and $t_n$ is the transmission coefficient of the periodic array of meta-atoms. Therefore, the efficiency of a low-NA metasurface is given by 
\begin{equation}
\centering
\eta_0= \frac{1}{4\pi^2}\left|\int_0^{2\pi}{t(\phi)e^{j\phi}\mathrm{d}\phi}\right|^2,
\end{equation}
where $t(\phi)$ is the complex-valued transmission coefficient of a periodic array composed of the meta-atom used for achieving the phase shift $\phi$.   We note that the effect of infrequent discontinuities violating the adiabatic metasurface approximation by gratings, which are caused by wrapping of $\phi_\mathrm{g}$, is ignored in efficiency estimations using (1) and (2).

\begin{figure}[t!]
\centering
\includegraphics[width=0.8\columnwidth]{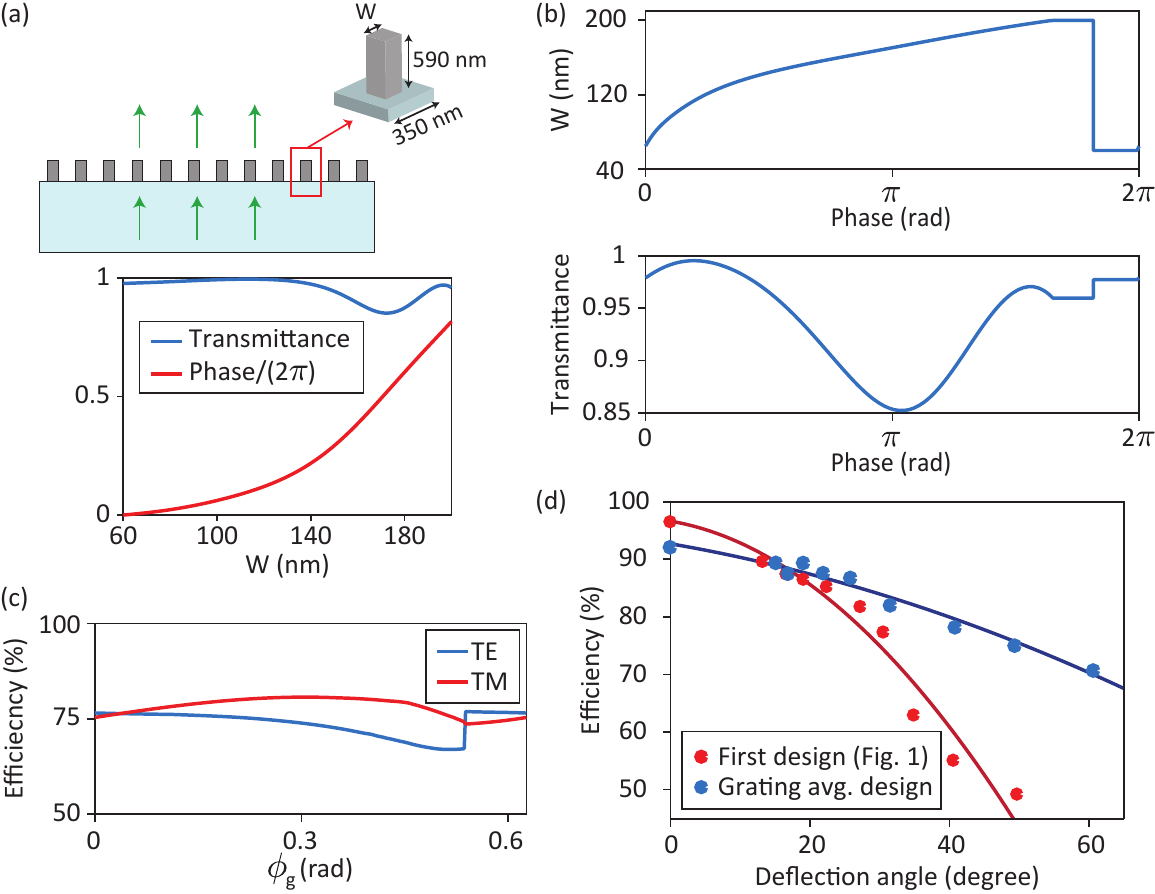}
\caption{(a) Schematic of a periodic metasurface based on the grating averaging design. Simulated transmission data for different nano-post widths $W$ is also shown. (b) Design curve relating the desired phase to the nano-post width for the metasurface design shown in (a) and the corresponding transmittance. The design curve is obtained using the transmission data shown in (a). (c)  Diffraction efficiencies of the third transmitted order of blazed gratings with different phases ($\phi_\mathrm{g}$). The blazed gratings are designed using the metasurface platform and design curve shown in (a) and (b). The gratings' period is 3500 nm (i.e., ten times the metasurface lattice constant) corresponding to a third order diffraction angle of 51.7$^\circ$. The efficiency values are computed for light normally incident from the substrate with transverse electric (TE) and  transverse magnetic (TM) polarizations. (d) Estimated deflection efficiencies of beam deflectors implemented using the metasurface design shown in Fig. 1 and the grating averaging design shown in (a) and (b).}
\end{figure}

\vspace{0.2in}
\noindent\textbf{Comparing different metasurface design platforms using the grating averaging technique}

The efficiency values obtained using the grating averaging technique can be used to evaluate and compare the  performance of different designs in implementing metasurfaces with different NAs. To illustrate the procedure, we consider a second design and compare its performance with the  design presented in Figs. 1a and 1b. The second design was selected to offer high efficiency at large deflection angles by exploring different designs and evaluating their efficiencies using (1).  A schematic of the second design is presented in Fig. 2a. In the second design, the nano-posts are 590 nm tall, the lattice constant is 350 nm, and the nano-post widths are varied between 60 nm and 200 nm. The transmittance and the phase of the transmission coefficient for 915 nm light normally incident on a periodic array of nano-posts with the parameters of this design are also shown in Fig. 2a. Nano-posts with widths larger than 200 nm are excluded in this design and the total phase shift covered by this design is $1.63\pi$ which is smaller than its ideal value of $2\pi$. The design curve that relates the nano-post width to the desired phase and the corresponding transmittance values for the second design are shown in Fig. 2b. Compared to the first design, the smaller phase shift coverage and the lower average transmittance of the second design indicate its inferior performance when used for implementing low-NA metasurfaces. In fact, the efficiency of low-NA metasurfaces designed using these designs can be obtained from (2) and the data presented in Figs. 1a and 2a, and are 96\% and 92\% for the first and second designs, respectively. However, high NA gratings based on the grating averaging design have higher efficiencies (Fig. 2c). 

\begin{figure}[t!]
\centering
\includegraphics[width=\columnwidth]{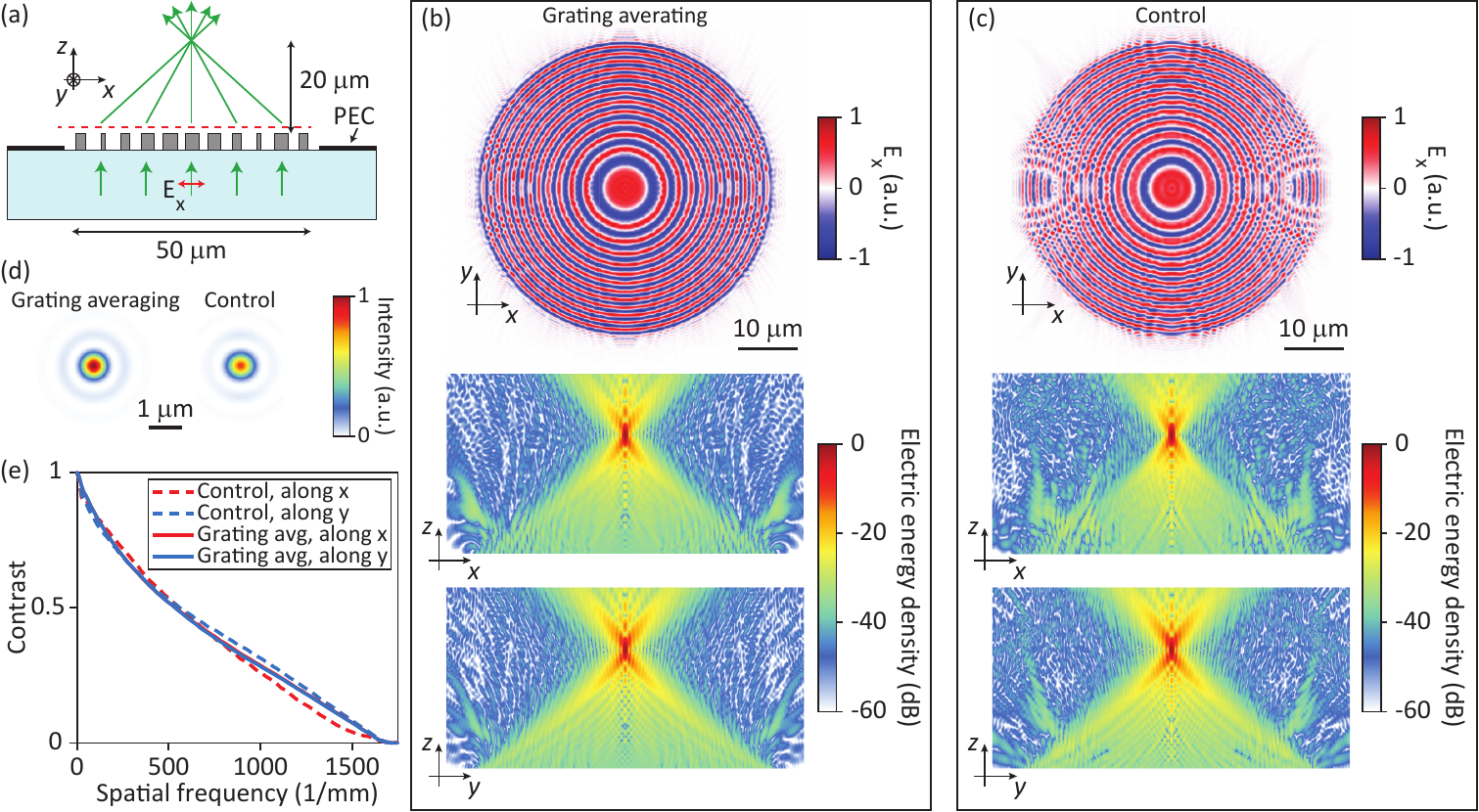}
\caption{(a) Schematic  of metalenses used in numerical simulations. Two metalenses are designed using the design curves shown Fig. 1b, referred to as the control metalens, and Fig.  2b which is referred to as the grating averaging metalens. (b) Full-wave simulation results of the grating averaging, and (c) Control metalenses. Top, $x$ components of the electric field over the output aperture of the metalenses. Bottom, logarithmic-scale electric energy density distributions on axial planes of the metalenses. (d) Focal plane intensity distributions for the grating averaging and control metalenses. (e) On-axis modulation transfer functions for the grating averaging and control metalenses. Simulations are performed at 915 nm. PEC: perfect electric conductor.}
\end{figure}

The diffraction efficiency of metasurfaces with higher NAs that are designed using these two designs can be computed using (1) by designing a set of gratings with different periods and phases $\phi_\mathrm{g}$. We selected different grating periods for each design and for each grating period we designed 40 gratings with different values of $\phi_\mathrm{g}$ (i.e., $\phi_\mathrm{g}=0$ to $2\pi$ in 40 steps). The diffraction coefficients of the gratings  ($t_n$) were found using RCWA simulations~\cite{Liu2012} and the deflection efficiencies of aperiodic beam deflectors were obtained by averaging the diffraction coefficients according to (1), finding the deflection efficiency as $\eta=|A|^2$, and averaging $\eta$ values for TE and TM polarizations. The unpolarized deflection efficiency versus deflection angle computed for these two design are shown in Fig. 2d. As Fig. 2d shows, for deflection angles larger than $\sim15^\circ$ the second design outperforms the first one. As mentioned earlier, the second design was found by exploring different designs and comparing their deflection efficiencies at large deflection angles using the grating averaging technique. Because the second design was found using the grating averaging technique, we  refer to it as the grating averaging design. We attribute the higher efficiency of the grating averaging design at large deflection angles to its smaller period and to the exclusion of nano-posts with large cross-sections. Such nano-posts support high order resonance modes that have off-axis nulls in their scattering patterns~\cite{Kamali2016,Kruk2017,Kuznetsov2016}.

\vspace{0.2in}
\noindent\textbf{Performance verification: Numerical results}

To demonstrate the effectiveness of the grating averaging technique in the design of more general metasurfaces, we designed a 50-$\mu$m-diameter metalens with a focal length of $f=20$ $\mu$m (NA of 0.78)  using the design curve presented in Fig. 2b. The phase profile of the metalens was chosen as $\Phi=-2\pi/\lambda_0\sqrt{x^2+y^2+f^2}$ to achieve aberration-free focusing for normally incident light with the vacuum wavelength of $\lambda_0=915$ nm. A schematic of the metalens is shown in Fig. 3a. The metalens is illuminated by an $x$-polarized normally incident plane wave and the light impinging outside the clear aperture of the metalens is blocked by a perfect electric conductor (PEC) as shown in Fig. 3a. The metalens was simulated using the finite difference time domain (FDTD) technique~\cite{Oskooi2010}. The $x$ component of the transmitted electric field on a plane a quarter of a wavelength above the top of the nano-posts, which is indicated by a dashed red line in Fig. 3a, is shown in Fig. 3b. The transmitted fields in the region above the metalens were computed via the plane wave expansion method~\cite{Born1999} and the electric energy density in the $xz$ and $yz$ cross-sections are shown in Fig. 3b. 

For comparison, we designed a metalens with similar parameters using the first design (Fig. 1a) and simulated it using the same procedure. The corresponding plots for this control metalens are presented in Fig. 3c. As $E_x$ field distributions presented in Figs. 3b and 3c show, the grating averaging metalens has a significantly smaller phase error at regions close to the circumference of the metalens where the deflection angles are larger. The low efficiency of the control metalens in deflecting the light toward the focal point means that some of the light is deflected to other directions. The interference of the light deflected to other directions and the light deflected toward the focus creates the phase error seen in Fig. 3c in areas close to the metalens circumference. The smaller phase error leads to a reduction in the light scattered to other directions as it can be seen in the logarithmic-scale energy distributions shown in Figs. 3b and 3c. The grating averaging metalens also has a higher transmission of 89\% compared with 75\% for the control metalens. Figure 3d shows the focal spots of the two metalenses. The grating averaging metalens has a more circular and brighter spot than the control metalens. 

The focusing efficiencies of the grating averaging and the control metalenses are found as 79\% and 63\%, respectively. The focusing efficiency is defined as the percentage of the power incident on the metalens aperture that is focused into and passes through a circle with a radius of 5 $\mu$m centered around the focal spot of the metalens. The radius of  5 $\mu$m for the aperture is selected for a direct comparison with the experimentally measured results that are presented in the next section. The on-axis modulation transfer function (MTF) of the two metalenses are shown in Fig. 3e. The control metalens MTF along the $x$ direction is smaller than the MTF of the grating averaging metalens at high spatial frequencies. The smaller MTF value is a result of the lower deflection efficiency of the control metalens along the $x$ axis and away from the center of the metalens where phase error is significant as it can be seen in the $E_x$ distribution shown in Fig. 3c.    

\vspace{0.2in}
\noindent\textbf{Performance verification: Experimental results}

To experimentally confirm the high efficiency of metasurfaces designed using the grating averaging technique, we designed and fabricated metalenses and metasurface beam deflectors. The metasurfaces were designed using the design curve shown in Fig. 2b and fabricated by depositing a 590-nm-thick layer of amorphous silicon on a fused silica substrate and pattering it using electron beam lithography and a dry etching process. Details of the fabrication process were similar to our previous work and can be found in Ref.~\cite{Arbabi2015}.

A schematic illustration of one of the fabricated metalenses is shown in Fig. 4a. The fabricated metalens  (diameter: 400 $\mu$m, focal length: 160 $\mu$m) is 8 times larger than the simulated metalens shown in Fig. 3a,  but has the same NA of 0.78. A scanning electron image of the fabricated device is shown in Fig. 4b. We measured the focal spot of the metalens using the measurement setup schematically shown in Fig. 4c. The metalens was illuminated by a collimated 915 nm laser beam and its focal plane intensity was magnified using the combination of the objective and tube lenses and captured by the camera (Fig. 4c). Figure 4d shows the measured focal spot for an $x$-polarized incident light. We did not observe any noticeable change in the focal spot as we varied the polarization of the incident beam. The measured focal spot intensity along the $x$-axis (indicated by the dashed black line) and the simulated focal spot intensity of an ideal metalens with a diameter of 400 $\mu$m and a focal length of 160 $\mu$m are also shown in Fig. 4d. The ideal metalens only modifies the optical wavefront of the incident light while keeping its local intensity unchanged. As Fig. 4d shows, the measured focal spot intensity matches well with the ideal metalens result, thus indicating a negligible wavefront aberration. The measured on-axis MTF of the fabricated metalens for $x$-polarized incident light and the MTF of an ideal metalens with the same NA are presented in Fig. 4e. The MTF of the measured device matches well with the ideal MTF at low spatial frequencies but drops faster at higher frequencies indicating the reduction of the deflection efficiency at larger deflection angles. 

\begin{figure}[t!]
\centering
\includegraphics[width=0.95\columnwidth]{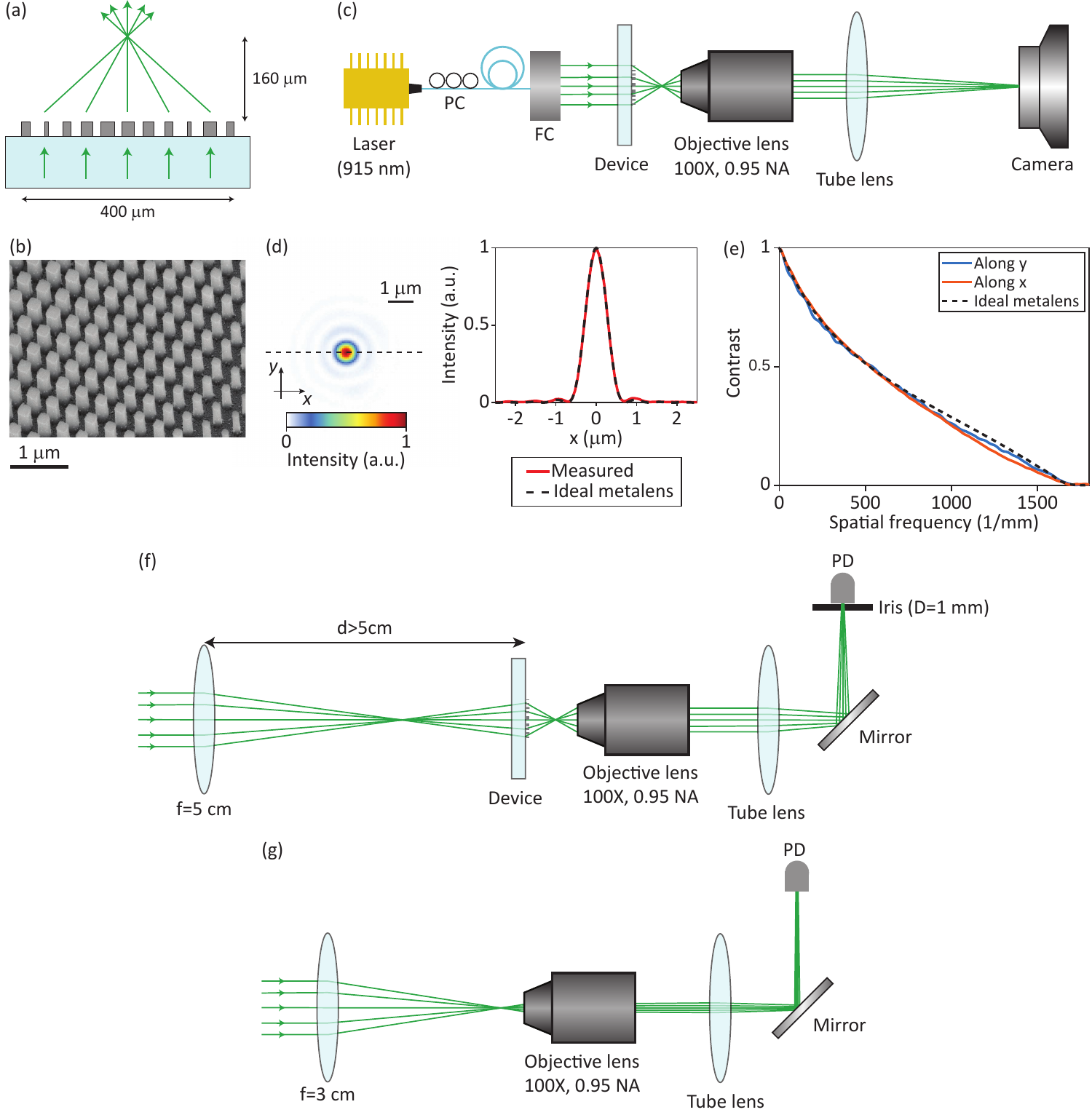}
\caption{(a) Illustration of the fabricated metalens. (b) A scanning electron image of a portion of the metalens. (c) Schematic of the measurement setup used for measuring the focal plane intensity distribution of the metalens. (d) The measured focal plane intensity distribution of the metalens (left) and measured focal plane intensity along the dashed black line shown in and the corresponding focal plane intensity data for an ideal metalens. (e) Measured on-axis modulation transfer function of the grating averaging  and the ideal metalenses. (f) Schematic of the setup used for measuring the optical power focused by the metalens, and (g) for measuring the incident optical power.  PC: polarization controller, FC: fiber collimator, PD: photodetector.}
\end{figure}

We measured the focusing efficiency of the metalens using the setup shown in Fig. 4f. The metalens was illuminated by a weakly diverging Gaussian beam that was generated by gently focusing the incident laser beam before the devices using  a lens with a focal length of 5 cm (as shown in Fig. 4f). The distance of the lens and the device was adjusted such that the beam radius at the device was  $\sim$140 $\mu$m thus more than 98\% of the incident power impinged on the device aperture (assuming a Gaussian intensity distribution). The intensity distribution at the metalens focal plane was magnified by 100$\times$, masked by passing through a 1-mm-diameter aperture in the image plane (corresponding to a 5-$\mu$m-radius aperture in the metalens focal plane),  and its power was measured. To measure the incident optical power, we focused the incident beam using a commercial lens (Thorlabs AC254-030-B-ML with a focal length of 3 cm and a transmission efficiency of 98\%) and measured its power in the image plane (Fig. 4g). The focusing efficiency of the metalens was found as 77\% by dividing the power passed through the aperture in Fig. 4f to the incident power.

In addition to the 400-$\mu$m-diameter metalens, using the grating averaging design shown in Figs. 2a and 2b,  we designed and fabricated 9 beam deflectors with different deflection angles ranging from 7$^\circ$ to 70$^\circ$, and a metalens with a diameter of 2 mm and a focal length of 800 $\mu$m.  The beam deflectors had a diameter of 400 $\mu$m and were illuminated by a normally incident Gaussian beam with a beam radius of  $\sim$100 $\mu$m on the device (Fig. 5a). The deflection efficiencies of the beam deflectors,  defined as the ratio of the deflected beam power to the incident power,  were measured for TE and TM polarizations. The average of the measured TE and TM deflection efficiencies (i.e., the deflection efficiency for unpolarized light) for different beam deflectors are presented in Fig. 5c. 

\begin{figure}[t!]
\centering
\includegraphics[width=0.7\columnwidth]{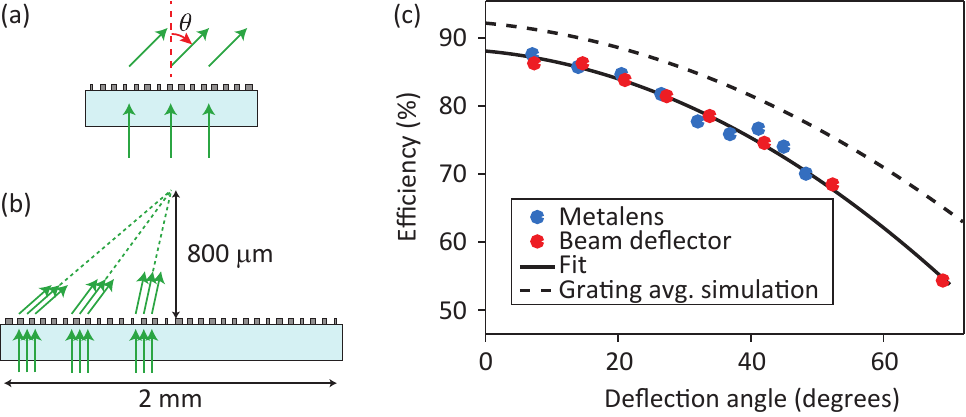}
\caption{(a) Schematic of the fabricated beam deflector, and (b) metalens that were used for determining the deflection efficiency as a function of deflection angle. The metalens was illuminated by a normally incident  Gaussian beam with a beam radius of $\sim$50 $\mu$m and the beam was scanned across the metalens aperture. (c) Measured deflection efficiency values for the beam deflectors and the metalens as functions of their deflection angles. The efficiency values are for unpolarized light and are measured at 915 nm. The solid  curve depicts a quadratic fit to the data, and the dashed curve is the grating averaging estimate that is also shown in Fig. 2d.}
\end{figure}

The 2-mm-diameter metalens was also used to evaluate the deflection efficiency at different locations over its aperture. As shown schematically in Fig. 5b,  we illuminated different regions of the metalens aperture with a normally incident  Gaussian beam which had a beam radius of $\sim$50~$\mu$m on the device. The transmitted beam power deflected toward the metalens focal point was measured for TE and TM polarizations and averaged to obtain the deflection efficiency for unpolarized light. The  measured efficiency values when the distance between the center of the incident beam and  the metalens axis was changed from  100 $\mu$m to 900 $\mu$m in 100 $\mu$m steps are shown in Fig. 5c  as a function of their corresponding deflection angles. As Fig. 5c shows, the local deflection efficiency of the metalens and the deflection efficiency of a beam deflector with the same deflection angle  are  almost equal, thus confirming the model of a metalens as a beam defector with a spatially varying deflection angle. Furthermore, the measured efficiencies follow the same trend as the simulated efficiencies obtained using the grating averaging technique (dashed line in Fig. 5c which is the fit to the simulated grating averaging data shown in Fig. 2c). The measured efficiencies are $\sim$7\% lower than their simulated values which can be attributed to the errors in the fabricated nano-posts' dimensions. The good agreement between the grating averaging  estimates and the measured values further confirms the accuracy of the proposed technique.

\section*{Discussion}
The grating averaging technique is a general approach for evaluating the performance of different metasurface designs. It can be used to estimate the efficiency of low- and high-NA metasurfaces and to compare different platforms. The technique relies on the idea of considering a gradient metasurface as a slowly varying grating and the realization that the diffraction efficiencies of different gratings with the same period designed using the same metasurface design curve can be vastly different. Although we illustrated the idea using the nano-post metasurfaces, the grating averaging technique is versatile and can be used in the design of a wide verity of metasurfaces composed of different meta-atom materials and geometries.

The evaluation of different designs can be performed quickly  because the grating diffraction coefficients can be computed in parallel and using fast simulation techniques such as RCWA. For example, the efficiency results presented in Fig. 2b were obtained using 40 cores on a workstation in less than 20 min. Once a high-performance design is found using the grating averaging technique, it can be used to design arbitrarily large metasurfaces. This is in contrast to computationally expensive optimization techniques that rely on full-wave simulations of the entire metasurface and inevitably limit the metasurface dimensions. Full-wave numerical simulations and experimental results of the fabricated metalenses and beam deflectors verify the accuracy and efficacy of the grating averaging technique. The grating averaging technique is expected to find widespread applications in the development of optical systems composed of cascaded  metasurfaces because the high efficiency is a crucial requirement in such systems.

\clearpage
\noindent\textbf{Funding}

This work was supported by Samsung Electronics. 

\noindent\textbf{Acknowledgements}

We gratefully acknowledge critical support and infrastructure provided for this work by the Kavli Nanoscience Institute at Caltech.

\noindent\textbf{Author contributions}

A.A. conceived the idea and designed the structures. A.A., E.A, S.M.K., and Y.H. fabricated the structures. A.A., M.M., and S.H. performed the simulations. A.A. and E.A. performed the measurements and analyzed the data. A.A. and A.F. devised the experiments and supervised the work. A.A. prepared the manuscript with input from all authors.

\newcommand{\BeginSupplementary}{
        \setcounter{figure}{0}
		\renewcommand{\figurename}{\textbf{Supplementary Figure}}
     }
\BeginSupplementary

\clearpage

\section*{Supplementary Note}
Here we show that the deflection coefficient of a beam deflector can be estimated using (1). Consider the aperiodic beam deflector  shown in Fig. 1e. Assume that the beam deflector is composed of a large number of extended cells and is illuminated with a normally incident plane wave with a power amplitude of 1. Depending on the polarization of the incident light, the deflected light is either TE or TM polarized.  The power amplitude of the deflected light $A$ along the deflection angle $\theta$ is related to the Fourier component of the electric or magnetic fields of the transmitted light on the output aperture of the device (dashed red line in Fig. 1e) at the spatial angular frequency of $k_0\sin(\theta)$, and can be found as~\cite{Harrington2001}
\begin{equation}
A=\frac{C}{L}\int_{0}^{L}{F(x)e^{jk_0\sin(\theta)x}\mathrm{d}x},
\end{equation} 
where $L$ is the length of the beam deflector. For TE-polarized incident light $F=E_y$  and $C=\sqrt{\frac{\cos(\theta)}{2Z}}$ is a constant that relates the electric field amplitude to the power amplitude, and $Z$ is the wave impedance in the $z>0$ region. For TM-polarized light $F=H_y$ and $C=\sqrt{2Z\cos(\theta)}$. The deflection efficiency is the square of the modulus of the power amplitude  and is given by $\eta=|A|^2$. Assuming that the beam deflector is composed of $M$ extended cells and each extended cell has a width of $\Lambda$, we can rewrite (3) as a sum of integrals over extended cells as
\begin{equation}
A=\frac{C}{L}\sum_{p=0}^{M-1}{\int_{p\Lambda}^{(p+1)\Lambda}{F(x)e^{jk_0\sin(\theta)x}\mathrm{d}x}}.
\end{equation} 
Assuming the beam deflector varies slowly from one extended cell to the next, fields over the $p^\mathrm{th}$ extended cell can be approximated  by the fields of a grating created by periodically repeating the same extended cell.  We denote the grating's transmitted field  over its output aperture by $F_{\mathrm{g}_p}(x)$. Thus,
\begin{equation}
A\approx\frac{C}{L}\sum_{p=0}^{M-1}{\int_{p\Lambda}^{(p+1)\Lambda}{F_{\mathrm{g}_{p}}(x)e^{jk_0\sin(\theta)x}\mathrm{d}x}}=\frac{C}{L}\sum_{p=0}^{M-1}{I_p},
\end{equation} 
where
\begin{equation}
I_p=\int_{p\Lambda}^{(p+1)\Lambda}{F_{\mathrm{g}_{p}}(x)e^{jk_0\sin(\theta)x}\mathrm{d}x}=\int_{p\Lambda}^{(p+1)\Lambda}{F_{\mathrm{g}_{p}}(x)e^{jk_0\sin(\theta_\mathrm{g})x}e^{jk_0(\sin(\theta)-\sin(\theta_\mathrm{g}))x}\mathrm{d}x}
\end{equation}
Note that the grating field and $e^{jk_0\sin(\theta_\mathrm{g})x}$ are periodic with a period of $\Lambda$, that is
\begin{equation}
 F_{\mathrm{g}_p}(x+\Lambda)e^{jk_0\sin(\theta_\mathrm{g})(x+\Lambda)}=F_{\mathrm{g}_p}(x)e^{jk_0\sin(\theta_\mathrm{g})x}. 
\end{equation}
Therefore, we can simplify (6) as
\begin{equation}
I_p=e^{jp\Delta\phi_\mathrm{g}}\int_{0}^{\Lambda}{F_{\mathrm{g}_{p}}(x)e^{jk_0\sin(\theta_\mathrm{g})x}e^{j\Delta\phi_\mathrm{g}\frac{x}{\Lambda}}\mathrm{d}x}\approx e^{jp\Delta\phi_\mathrm{g}}\int_{0}^{\Lambda}{F_{\mathrm{g}_{p}}(x)e^{jk_0\sin(\theta_\mathrm{g})x}\mathrm{d}x},
\end{equation}
where we have defined $\Delta\phi_\mathrm{g}=k_0(\sin(\theta)-\sin(\theta_\mathrm{g}))\Lambda$ and used the approximation $e^{j\Delta\phi_\mathrm{g}\frac{x}{\Lambda}}\approx1$ for $\Delta\phi_\mathrm{g}\ll1$. $\Delta\phi_\mathrm{g}$ represents the phase shift from one extended  cell to the next.  $I_p$ can be expressed in terms of the grating diffraction coefficients as
\begin{equation}
I_p\approx e^{jp\Delta\phi_\mathrm{g}}\int_{0}^{\Lambda}{F_{\mathrm{g}_{p}}(x)e^{jk_0\sin(\theta_\mathrm{g})x}\mathrm{d}x}=\frac{\Lambda}{C} e^{jp\Delta\phi_\mathrm{g}} t_n(p\Delta\phi_\mathrm{g}),
\end{equation} 
where
\begin{equation}
 t_n(p\Delta\phi_\mathrm{g})=\frac{C}{\Lambda}\int_{0}^{\Lambda}{F_{\mathrm{g}_{p}}(x)e^{jk_0\sin(\theta_\mathrm{g})x}\mathrm{d}x},
\end{equation} 
represents the diffraction coefficient of the $n^\mathrm{th}$ diffraction order of an $n^\mathrm{th}$-order blazed grating that is designed with the phase of $p\Delta\phi_\mathrm{g}$. Plugging $I_p$ from (9) into (5) we obtain
\begin{equation}
A\approx\frac{\Lambda}{D}\sum_{p=0}^{M-1}{t_n(p\Delta\phi_\mathrm{g})e^{jp\Delta\phi_\mathrm{g}}}=\frac{1}{M}\sum_{p=0}^{M-1}{t_n(p\Delta\phi_\mathrm{g})e^{jp\Delta\phi_\mathrm{g}}},
\end{equation} 
which is the average of $t_{n}(\phi_\mathrm{g})e^{j\phi_\mathrm{g}}$ over different extended cells. Because $\Delta\phi_\mathrm{g}\ll1$, $M\gg1$, and $t_{n}(\phi_\mathrm{g})e^{j\phi_\mathrm{g}}$ is periodic with a period of $2\pi$,  its average can also be computed as an integral over its period as
\begin{equation}
A\approx\frac{1}{2\pi}\int_0^{2\pi}{t_{n}(\phi_\mathrm{g})e^{j\phi_\mathrm{g}}\mathrm{d}\phi_\mathrm{g}},
\end{equation}
which is the result presented in (1).

\end{document}